%% file: QCRF_arXiv_v3.tex
\ifdefined\manu
\documentclass[12pt]{article}
\else
\documentclass[twocolumn,10pt]{article} 
\fi

\usepackage[square,numbers,sort&compress,comma]{natbib}

\usepackage{amsmath}
\usepackage{amssymb}
\usepackage{caption}
\usepackage{graphicx}
\usepackage{latexsym}
\usepackage{times}
\usepackage[pagewise]{lineno}

\ifdefined\manu
\usepackage{geometry}
\else
\topmargin - 12pt 
\fi

\renewenvironment{abstract}%
              {
               \small
               {\bfseries \abstractname}
               \par
               \vspace{10pt}
              }

\renewcommand\abstractname{Abstract}

\newcommand{\nomenclature}
              [1]
              {
               \bgroup
               \flushleft
               \small\bf
               #1
               \par
               \egroup
              }

\renewcommand{\section}
              [1]
              {
               \bgroup
               \flushleft
               \small\bf
               \refstepcounter{section}
               \arabic{section}. #1
               \par
               \egroup
              }

\renewcommand{\subsection}
              [1]
              {
               \bgroup
               \flushleft
               \small\em
               \refstepcounter{subsection}
               \arabic{section}.
               \arabic{subsection}. #1
               \par
               \egroup
              }

\renewcommand{\subsubsection}
              [1]
              {
               \bgroup
               \flushleft
               \small\em
               \refstepcounter{subsubsection}
               \arabic{section}.
               \arabic{subsection}.
               \arabic{subsubsection}. #1
               \par
               \egroup
              }

  \newcommand{\acknowledgement}
              [1]
              {
               \bgroup
               \flushleft
               \small\bf
               #1
               \par
               \egroup
              }

  \newcommand{\sectionbib}
              [1]
              {
               \bgroup
               \flushleft
               \small\bf
               #1
               \par
               \egroup
              }

\usepackage{setspace}
\usepackage{bm}
\usepackage{booktabs}
\usepackage{multirow}
\usepackage{algorithm}
\usepackage{algpseudocode}
\usepackage{physics}
\usepackage{qcircuit}
\usepackage{xcolor}
\usepackage{pgf}

\graphicspath{{Fig/}}

\newcommand{\EQ}{\begin{equation}}
\newcommand{\EN}{\end{equation}}
\newcommand{\SEQ}{\begin{subequations}}
\newcommand{\SEN}{\end{subequations}}
\newcommand{\EQA}{\begin{eqnarray}}
\newcommand{\ENA}{\end{eqnarray}}
\newcommand{\CS}{\begin{dcases}}
\newcommand{\CN}{\end{dcases}}
\newcommand{\AR}{\begin{array}}
\newcommand{\AN}{\end{array}}

\newcommand{\mc}{\mathcal}
\newcommand{\mr}{\mathrm}
\newcommand{\mbb}{\mathbb}

\newcommand{\bv}{\bm{v}}
\newcommand{\bw}{\bm{w}}
\newcommand{\bA}{\bm{A}}
\newcommand{\bD}{\bm{D}}
\newcommand{\bH}{\bm{H}}
\newcommand{\bI}{\bm{I}}
\newcommand{\bphi}{\bm{\phi}}

\newcommand{\etal}{\textit{et al}.}

\newcommand{\lrr}[1]{\left(#1\right)}
\newcommand{\lrs}[1]{\left[#1\right]}

\newcommand{\lrn}[1]{\left\vert#1\right\vert}
\newcommand{\lrN}[1]{\left\Vert#1\right\Vert}

\begin{document}

\title{\LARGE
Quantum computing of reacting flows via Hamiltonian simulation
}

\author{{\large Zhen Lu$^{a}$,
                Yue Yang$^{a,b,*}$
        }\\[10pt]
        {\footnotesize \em
            $^a$State Key Laboratory for Turbulence and Complex Systems, College of Engineering,
        }\\[-5pt]
        {\footnotesize \em
            Peking University, Beijing 100871, China
        }\\[-5pt]
        {\footnotesize \em
            $^b$HEDPS-CAPT, Peking University, Beijing 100871, China
        }}

\date{}


\small
\baselineskip 10pt


\ifdefined\manu
\else
\twocolumn[\begin{@twocolumnfalse}
\fi

\vspace{50pt}
\maketitle
\vspace{40pt}
\rule{\textwidth}{0.5pt}
\begin{abstract} 
We report the quantum computing of reacting flows by simulating the Hamiltonian dynamics.
The scalar transport equation for reacting flows is transformed into a Hamiltonian system, mapping the dissipative and non-Hermitian problem in physical space to a Hermitian one in a higher-dimensional space.
Using this approach, we develop the quantum spectral and finite difference methods for simulating reacting flows in periodic and general conditions, respectively.
The present quantum computing algorithms offer a ``one-shot'' solution for a given time without temporal discretization, avoiding iterative quantum state preparation and measurement.
We compare computational complexities of the quantum and classical algorithms.
The quantum spectral method exhibits exponential acceleration relative to its classical counterpart,
and the quantum finite difference method can achieve exponential speedup in high-dimensional problems.
The quantum algorithms are validated on quantum computing simulators with the Qiskit package.
The validation cases cover one- and two-dimensional reacting flows with a linear source term and periodic or inlet-outlet boundary conditions.
The results obtained from the quantum spectral and finite difference methods agree with analytical and classical simulation results.
They accurately capture the convection, diffusion, and reaction processes.
This demonstrates the potential of quantum computing as an efficient tool for the simulation of reactive flows in combustion.
\end{abstract}
\vspace{10pt}
\parbox{1.0\textwidth}
{\footnotesize {\em Keywords:}
Quantum Computing;
Reacting Flow;
Scalar Transport;
Hamiltonian Simulation
}
\rule{\textwidth}{0.5pt}
\vspace{10pt}

\ifdefined\manu
\else
\end{@twocolumnfalse}]
\fi

\clearpage


\ifdefined\manu
\doublespacing
\fi


\section{Introduction\label{sec:intro}} \addvspace{10pt}

Quantum computing has gained widespread attention as a cutting-edge area of modern science~\cite{Nielson2010}.
Quantum computers offer the potential to solve certain problems much more efficiently than classical computers by leveraging quantum physics, promising groundbreaking applications including the potential to reduce computational cost and carbon footprint.

Simulation of quantum dynamics, also referred to as Hamiltonian simulation, is a crucial application of quantum computing~\cite{Feynman1982}, and has achieved significant advances in recent decades~\cite{Lloyd1996,Low2017,Childs2018,Campbell2019}.
Furthermore, quantum computing can be utilized in simulating classical systems, including reacting flows~\cite{Givi2020}.
For instance, Xu \etal~\cite{Xu2019} calculated the reactant conversion rate in partially stirred reactors using quantum metrology.
Akiba \etal~\cite{Akiba2023} integrated a chemical source term in quantum computing via the Carleman linearization.

The reacting flows are governed by partial differential equations (PDEs).
Although the development in high-performance computing has made the direct numerical simulation (DNS)~\cite{Domingo2023} as a useful tool, broadening the scope on parameters and geometries in turbulent combustion research~\cite{Lu2021,Su2022},
the demanding computational expense remains a major obstacle in performing large-scale DNS.
The integration of computational fluid dynamics with quantum computing can be promising for developing next-generation simulation methods.

There has been a growing interest in solving PDEs using quantum algorithms.
Upon spatial and temporal discretizations, PDEs can be transformed into a set of linear algebraic equations.
The quantum algorithm speedup in solving these equations~\cite{Harrow2009} can lead to the potential acceleration~\cite{Childs2021}. 
Clader \etal~\cite{Clader2013} extended the Harrow-Hassidim-Lloyd algorithm~\cite{Harrow2009} for solving PDEs by first discretizing, and then reducing the problem to solving linear systems.
Subsequently, additional methods were proposed~\cite{Childs2021,Krovi2023} to decrease computational complexity and apply to different problems~\cite{Costa2019,Wei2023,Liu2023}.
Nonetheless, due to time discretization and quantum collapse, iterative preparation and measurement of quantum states are necessary for simulating over a specified period, but the cost of input/output in quantum computing can be prohibitive~\cite{Hoefler2023}.

An intuitive way to avoid this issue is to transform the flow governing equation to a Schr\"odinger equation.
Meng and Yang~\cite{Meng2023,Yang2023,Meng2024a} proposed a framework for quantum computing of fluid dynamics based on the hydrodynamic Schr\"odinger equation (HSE), by generalizing the Madelung transform.
Time marching of the HSE is expressed as unitary operators, making it efficient for quantum computing.
Similarly, if the spatial discretization of PDEs yields a system of ordinary differential equations (ODEs) as a Hamiltonian system, the solution is accessible through Hamiltonian simulation efficiently~\cite{An2023}.
On the other hand, finding a mapping of the transport equation in reacting flows to a Hamiltonian system can be challenging due to the dissipative diffusion and nonlinear reaction terms.

Recently, Jin \etal~developed a theory of Schr\"odingerisation~\cite{Jin2023c} to convert a linear PDE to a series of independent Schr\"odinger equations.
We apply the Schr\"odingerisation to quantum computing of reacting flows, and develop quantum spectral and finite difference methods.
By implementing the quantum algorithms via the quantum circuit and estimating computational complexities, we demonstrate the applicability of quantum computing of several reacting flows with inlet-outlet or periodic boundaries.

Note that the current quantum computing is in the noisy intermidiate-scale quantum era~\cite{Preskill2018,Daley2022}. 
Imperfect control of qubits introduces noise, and the number of available qubits is limited. 
Consequently, we conducted the quantum simulation using the Qiskit~\cite{Qiskit}, a quantum computing simulator, on a classical computer at this stage.


\section{Quantum computing via Hamiltonian simulation\label{sec:theory}} \addvspace{10pt}

Quantum computing is motivated to efficiently simulate the Hamiltonian dynamics~\cite{Feynman1982,Lloyd1996}
\EQ\label{eq:Hamiltonian}
    i\dv{\psi}{t} = H\psi, \quad \psi\lrr{t=0} = \psi_0,
\EN
where $i=\sqrt{-1}$, $\psi\lrr{t,x}$ is a wave function of time $t$ and space $x$, which can be expressed as a quantum state $\ket{\psi}$, and the Hamiltonian $H$ is Hermitian, i.e., $H=H^\dagger$, where $\dagger$ denotes conjugate transpose.
Note that Eq.~\eqref{eq:Hamiltonian} describes a linear system without dissipation.
If $H$ is time-independent, Eq.~\eqref{eq:Hamiltonian} has the solution $\psi\lrr{t}=\exp\lrr{-iHt}\psi\lrr{0}$.
Subsequently, a quantum circuit can be implemented to produce the unitary operator $U=\exp\lrr{-iHt}$ for the evolution of the quantum state
\EQ\label{eq:evo}
    \ket{\psi\lrr{t}} = U\ket{\psi\lrr{0}}.
\EN

In reacting flows, the transport of a scalar $\phi$ is governed by
\EQ\label{eq:transport}
    \rho\pdv{\phi}{t} + \rho u \pdv{\phi}{x} = \pdv{}{x}\lrr{\rho D \pdv{\phi}{x}} + S\lrr{\phi},
\EN
where $\rho$ is the density, $u$ the velocity, $D$ the diffusivity, and $S$ the reaction source term.
For simplicity, we assume a constant density $\rho=1$, constant diffusivity, and linear reaction source term $S=\alpha \phi$.
Thus, Eq.~\eqref{eq:transport} becomes
\EQ\label{eq:linearPDE}
    \pdv{\phi}{t} + u \pdv{\phi}{x} = D\pdv[2]{\phi}{x} + \alpha \phi.
\EN
Here we only present the algorithm in 1D and 2D flows, but it is straightforward to extend it to multi-dimensional problems.

Although the reacting flow is generally characterized by a nonlinear source term, quantum computing of the linear PDE is still a meaningful endeavor.
There are ongoing efforts to develop various linearization methods to apply quantum computing to classical nonlinear problems~\cite{Liu2021,Akiba2023}.
Alternatively, nonlinear dynamics can be mapped to a linear representation through the Liouville equation or the Koopman-von Neumann approach~\cite{Joseph2020,Jin2023a}.

To solve Eq.~\eqref{eq:linearPDE} by the Hamiltonian simulation in Eq.~\eqref{eq:Hamiltonian}, it is necessary to convert the dissipative equation to a non-dissipative one.
We adopt the warped phase transform~\cite{Jin2023c}
\EQ\label{eq:warp}
    w\lrr{t,x,p} = e^{-p}\phi\lrr{t,x}, \; p \geq 0,
\EN
which introduces an auxiliary variable $p$.

\subsection{Spectral method}\label{subsec:spectral} \addvspace{10pt}

For a special case with periodic boundary conditions and a constant velocity $u=c$, Eq.~\eqref{eq:linearPDE} can be solved using the spectral method.
Substituting Eq.~\eqref{eq:warp} into \eqref{eq:linearPDE} yields
\EQ\label{eq:wPeriodic}
    \pdv{w}{t} + c\pdv{w}{x} = -D\pdv{}{p}\pdv[2]{w}{x} - \alpha\pdv{w}{p}, \; p \geq 0.
\EN
The Fourier transform of Eq.~\eqref{eq:wPeriodic} on $x$ reveals a linear convection in the $p$-dimension at a speed of
\EQ\label{eq:speedSP}
a = -D\zeta^2+\alpha,
\EN
where $\zeta$ is the wavenumber of $x$.
If $a \leq 0$, we can extend Eq.~\eqref{eq:wPeriodic} to $p < 0$ using the initial profile $ w\lrr{0,x,p} = e^{-\lrn{p}} \phi \lrr{0,x} $.
Applying the Fourier transform of $w$ on $p$, i.e., $\psi\lrr{t,x,\eta}=\mc{F}_p\lrs{w}\lrr{\eta}$, we have
\EQ\label{eq:periodic}
    \pdv{\psi}{t} + c\pdv{\psi}{x} = -i \eta D\pdv[2]{\psi}{x}-i\eta\alpha\psi.
\EN
Note that the warped phase transform makes the dissipative, non-Hermitian diffusion and reaction terms in the $x$ space Hermitian in the $x$-$\eta$ space.
Applying the Fourier transform of $\psi$ on $x$, $\hat{\psi}\lrr{t,\zeta,\eta} = \mc{F}_x\lrs{\psi\lrr{t,x,\eta}}\lrr{\zeta}$, gives
\EQ\label{eq:spectral}
    i\dv{\hat{\psi}}{t} = \lrr{ c \zeta - D \eta \zeta^2 + \alpha \eta}\hat{\psi},
\EN
which corresponds to Eq.~\eqref{eq:Hamiltonian} with $H=c \zeta - D \eta \zeta^2 + \alpha\eta$ for given $\zeta$ and $\eta$.

Equation~\eqref{eq:spectral} enables us to solve the problem via the Hamiltonian simulation.
Discretizing Eq.~\eqref{eq:spectral} in $\zeta$ and $\eta$ with $N_x$ and $N_p$ points, respectively, yields
\EQ\label{eq:spectralMatrix}
    i\dv{\hat{\bm{\Psi}}}{t} \!\!=\!\! \lrr{
        c \bD_\zeta \!\otimes\! \bI_p
        \!\!-\!\! D \bD_\zeta^2 \!\otimes\! \bD_\eta
        \!\!+\!\! \alpha \bI_x \!\otimes\! \bD_\eta
    }\! \hat{\bm{\Psi}},
\EN
where $\hat{\bm{\Psi}}\lrr{t}$ is a $N_x N_p$-dimensional vector with $\hat{\Psi}_{k+\lrr{j-1}N_p}=\hat{\psi}\lrr{t,\zeta_j,\eta_k}$, $j=1,\cdots,N_x$, $k=1,\cdots,N_p$;
$\bD_\zeta=\mr{diag}\lrr{\zeta_1,\cdots,\zeta_{N_p}}$ and $\bD_\eta=\mr{diag}\lrr{\eta_1,\cdots,\eta_{N_p}}$ are diagonal matrices with entries of the wavenumbers $\zeta_j$ and $\eta_k$, respectively;
$\bI_x$ and $\bI_p$ are $N_x\times N_x$ and $N_p\times N_p$ identity matrices, respectively.
Using Eq.~\eqref{eq:evo} to obtain $\hat{\psi}\lrr{t,\zeta,\eta}$, we obtain $w\lrr{t,x,p}$ through the inverse Fourier transform on $x$ and $p$.
For $p \geq 0$, we have the solution $\phi\lrr{t,x} = e^{p}w\lrr{t, x, p}$.

\subsection{Finite difference method}\label{subsec:discretized} \addvspace{10pt}

For a general case with periodic, inlet, or outlet boundaries, we discretize Eq.~\eqref{eq:linearPDE} in $x$ using the finite difference method on $N_x$ spatial points, resulting the ODEs
\EQ\label{eq:ODE}
    \dv{\bphi}{t} = \bA\bphi + \bm{b},
\EN
where $\bphi$ and $\bm b$ are $N_x$-dimensional vectors, with $\bphi_j\lrr{t}=\phi\lrr{t,x_j}$, and $\bA$ is a $N_x\times N_x$ matrix.
It suffices to assume $\bm{b}=\bm{0}$. Otherwise, using an auxiliary variable $\theta\lrr{t}=1$ can convert the inhomogeneous ODEs to homogeneous ones via
\EQ
    \bphi' = \begin{bmatrix} \bphi \\ \theta \end{bmatrix}
    \;\mr{and}\;
    \bA' = \begin{bmatrix} \bA & \bm{b} \\ \bm{0} & 0 \end{bmatrix}.
\EN

The matrix $\bA$ is generally neither anti-Hermitian nor Hermitian, so Eq.~\eqref{eq:ODE} cannot be solved by the Hamiltonian simulation in Eq.~\eqref{eq:evo} directly.
Meanwhile, $\bA$ can be decomposed as
\EQ
    \bA = \bH_1 + i \bH_2,
\EN
where both $\bH_1=\lrr{\bA+\bA^\dagger}/2$ and $\bH_2=\lrr{\bA-\bA^\dagger}/2i$ are Hermitian.

Applying the warped phase transform in Eq.~\eqref{eq:warp} yields
\EQA
    \pdv{\bw}{t} = -\bH_1\pdv{\bw}{p} + i\bH_2 \bw, \label{eq:wDiscretized} \\
    \bw\lrr{0,p} = e^{-\lrn{p}}\bphi\lrr{0}.
\ENA
Like Eq.~\eqref{eq:wPeriodic}, Eq.~\eqref{eq:wDiscretized} is a series of linear convection equations in the $p$-dimension with a speed of $a=\lambda_j$ for the $j$-th eigenvector of $\bH_1$ with its eigenvalue $\lambda_j$. 
To ensure the extension of Eq.~\eqref{eq:wDiscretized} to $p<0$, it requires that $\lambda_j \leq 0$, i.e., $\bH_1$ should be negative semi-definite.
Applying the Fourier transformation of Eq.~\eqref{eq:wDiscretized} on $p$, we obtain a series of independent ODEs
\EQ\label{eq:one}
    i \dv{\bm{\psi}}{t} = \lrr{ \eta \bH_1 - \bH_2 } \bm{\psi}
\EN
for each wavenumber $\eta$ in $p$.
Equation~\eqref{eq:one} describes the Hamiltonian dynamics where $\bH_\eta=\eta\bH_1 + \bH_2$ is Hermitian.

Discretizing in $p$ using $N_p$ points, Eq.~\eqref{eq:one} can be expressed as
\EQ\label{eq:full}
    i \dv{\bm{\Psi}}{t} = \lrr{ \bH_1 \otimes \bD_\eta - \bH_2 \otimes \bI_p } \bm{\Psi}.
\EN
Upon obtaining the solution $\bm\Psi\lrr{t}$, $\bw\lrr{t,p}$ can be obtained by the inverse Fourier transform of $\bm\Psi$ on $p$.
Then, $\bphi\lrr{t}=e^{p}\bw\lrr{t,p}$ for $p\geq 0$.

Note that Eqs.~\eqref{eq:spectralMatrix} and \eqref{eq:full} have no discretization in $t$.
Accordingly, the Schr\"odingerisation technique offers a ``one-shot'' solution, avoiding iterative time marching with multiple time steps~\cite{Jin2023c}.
This feature is crucial for quantum computing, because there is no need of costly quantum state measurements and preparations at intermediate time steps.
A detailed derivation can be found in supplementary material.

\section{Quantum computing implementation\label{sec:implement}} \addvspace{10pt}

\subsection{Quantum algorithms} \addvspace{10pt}

We discretize $w$ in $x$ and $p$ over $\lrs{-L_x/2, L_x/2}$ and $\lrs{-L_p/2, L_p/2}$ using $N_x$ and $N_p$ points, with mesh sizes $\Delta x = L_x/N_x$ and $\Delta p = L_p / N_p$, respectively.
For brevity, we assume the ODEs obtained via discretization are homogeneous.
In the inhomogeneous case, $N_x-1$ points are used to discretize in $x$
with another point for $\theta\lrr{t}=1$.

We use $n_x = \log_2 \lrr{N_x} $ qubits to encode the variable $\phi\lrr{t,x}$, with the state vectors $\ket*{j_1}, \cdots, \ket*{j_{n_x}}$.
Each qubit in $\ket*{j_l}, l=1,\cdots,n_x$ has the computational basis state $\ket*{0}$ or $\ket*{1}$.
The computational basis $\ket*{j}=\ket*{j_1 j_2 \cdots j_{n_x}}$ then represents the computational domain in the Hilbert space $\mbb{C}^{2^{n_x}}$, with $j=\sum_{l=1}^{n_x}j_{l}2^{n_x-l}$~\cite{Nielson2010,Meng2023} and $\ket*{j_1j_2\cdots j_{n_x}} \equiv \ket*{j_1}\otimes\ket*{j_2}\otimes\cdots\otimes\ket*{j_{n_x}}$.
Similar to $x$, we employ a set of computational basis $\ket*{k}=\ket*{k_1k_2\cdots k_{n_p}}$ using $n_p=\log_2\lrr{N_p}$ qubits for the $p$-dimension.

In this way, we encode the quantum state as
\EQ
    \ket*{\phi\lrr{t}} = \dfrac{1}{\lrN{\bphi}}\sum_{j=1}^{N_x} \phi\lrr{t,x_j} \ket*{j},
\EN
where $x_j = -L_x/2 + \lrr{j-1}\Delta x$,
$\lrN{\cdot}$ denotes the $L_2$ norm.
%
Together with the quantum state
\EQ
    \ket*{v} = \dfrac{1}{\lrN{\bv}}\sum_{k=1}^{N_p} v\lrr{p_k} \ket*{k}
\EN
with $v\lrr{p} = \exp\lrr{-\lrn{p}}$
and $p_k = -L_p/2 + \lrr{k-1}\Delta p$,
the quantum state for the initial solution $w\lrr{0,x,p}$ is
\EQ\label{eq:wState}
    \ket*{w\lrr{0}} = \ket*{\phi\lrr{0}} \otimes \ket*{v}
\EN
with $n_x+n_p$ qubits.

Using the spectral method, the solution is
\EQ\label{eq:evoSP}
\begin{split}
    \ket*{w\lrr{t}} =
    & (\mc{QF}_x^\dagger\otimes\mc{QF}_p^\dagger)\exp\lrr{-i\bH t} \\
    & \times(\mc{QF}_x\otimes\mc{QF}_p)\ket*{w\lrr{0}},
\end{split}
\EN
where $\bH = c \bD_\zeta \!\otimes\! \bI_p \!-\! D \bD_\zeta^2 \!\otimes\! \bD_\eta \!+\! \alpha \bI_x \!\otimes\! \bD_\eta $ corresponds to Eq.~\eqref{eq:spectralMatrix},
and $\mc{QF}$ denotes the quantum Fourier transform (QFT)~\cite{Weinstein2001}.

Using the finite difference method, the solution $\ket*{w\lrr{t}}$ is obtained as
\EQ\label{eq:evoFD}
\begin{split}
    \ket*{w\lrr{t}} =
    & (\bI_x\otimes\mc{QF}_p^\dagger)\exp\lrr{-i\bH t} \\
    & \times(\bI_x\otimes\mc{QF}_p)\ket*{w\lrr{0}},
\end{split}
\EN
where $\bH = \bH_1 \otimes \bD_\eta - \bH_2 \otimes \bI_p $ corresponding to Eq.~\eqref{eq:full}.
In the following, we will refer to the methods for Eqs.~\eqref{eq:evoSP} and \eqref{eq:evoFD} as quantum spectral (SP) and quantum finite difference (FD) methods, respectively.

Since there is no temporal discretization, the solution at a given time is obtained by ``one shot''.
Due to the collapse of a quantum state on measurement, frequent measurement of the quantum state is costly and should be avoided.
The present quantum state preparation and measurement apply only at the beginning and end of the simulation, respectively.
The maximum simulation time is determined by the $p$-dimension, $L_p/(2\lrN{\bm{a}}_\infty)$, where $\lrN{\cdot}_\infty$ denotes the $L$-$\infty$ norm.
In addition to the influence on the maximum simulation time, the value of $\Delta p=L_p/N_p$ also plays a significant role in determining the accuracy of the solution.

\subsection{Quantum circuit\label{subsec:implement}} \addvspace{10pt}

\begin{figure}[h!]
    \centerline{
        \scalebox{0.75}{
            \input{Fig/figQuantumCircuitSpectralFull.tex}
        }
    }
    \caption{
        Overall quantum circuit for the quantum SP method.
    }
    \label{fig:qc}
\end{figure}
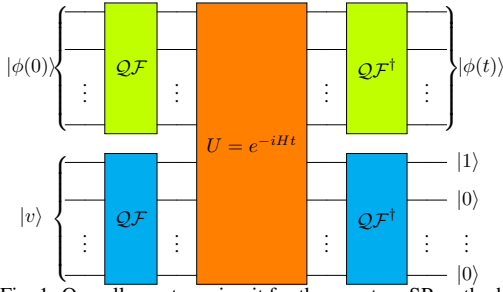

A quantum algorithm is typically represented by the quantum circuit.
Figure~\ref{fig:qc} displays the overall quantum circuit implementing the quantum SP method to compute $\phi\lrr{t}$.
For the quantum FD method, the corresponding quantum circuit simply does not contain $\mc{QF}_x$ and $\mc{QF}_x^\dagger$ on the first $n_x$ qubits.
Note that $U$ changes with the Hamiltonian in every implementation.
When the last $n_p$ qubits of $\ket*{w\lrr{t}}$ are conditioned on the computational basis $\ket*{k}$ with $k\geq N_p/2$, the first $n_x$ qubits store $e^{-p_k}\ket*{\phi\lrr{t}}$, where $\phi\lrr{t,x}$ is the desired solution of Eq.~\eqref{eq:linearPDE}.
For example, conditioning the last $n_p$ qubits on $\ket*{10\cdots 0}$, corresponding to $k=N_p/2$ and $p_k=0$, the first $n_x$ qubits store $\ket*{\phi\lrr{t}}$.


Due to the limitation on the number of available qubits~\cite{Nielson2010} and the efficiency of quantum computing simulators~\cite{Qiskit}, it is preferred to validate the quantum algorithm on a small number of qubits.
For a wavenumber $\eta$ in the auxiliary dimension, Eqs.~\eqref{eq:spectral} and \eqref{eq:one} correspond to a set of general Schr\"odinger equations; each set is independent.
Therefore, we performed the Fourier (or inverse) transform on $p$ in a pre (or post)-processing way on a classical computer.
This treatment is only for simulating the flow with more qubits based on limited computational resources, though the present algorithm can be fully conducted on a quantum computer.

On the detailed implementation, the QFT can be implemented by $\mc{O}\lrr{n^2}$ quantum gates on $n$ qubits, achieving an exponential acceleration compared to $\mc{O}\lrr{n2^n}$ operations of the fast Fourier transform~\cite{Weinstein2001}.
%
%
An arbitrary unitary operator can be exactly expressed using single qubit and CNOT gates~\cite{Nielson2010}.
Particularly, the diagonal Hamiltonian in Eq.~\eqref{eq:spectralMatrix} of the quantum SP method can be performed efficiently with $\mathcal{O}\left(n^2\right)$ gates~\cite{Meng2023,Meng2024b}.
An example of the quantum circuit implementation of $U=\exp(-i\bm{D}_\zeta^2)$ on three qubits is illustrated in Fig.~\ref{fig:qcfold}, which consists of the Hadamard, CNOT, and general unitary gates~\cite{Qiskit}.
As for the quantum FD method, the complexity of performing the unitary operator in Eq.~\eqref{eq:full} is $\mc{O}\lrr{s\lrr{\bA}\lrN{\bA}_{\mr{max}}\lrr{n_x+n_p}2^{n_p}}$~\cite{Jin2023c}, where $s\lrr{\bA}$ denotes the sparsity of $\bA$ and $\lrN{\bA}_{\mr{max}}$ represents the largest entry of $\bA$ in absolute value.
We prepare the initial state $\ket{w\lrr{0}}$ by separated initializations of $\ket{\phi\lrr{0}}$ and $\ket{v}$, as in Eq.~\eqref{eq:wState}.
The initialization is performed through the Qiskit.

\ifdefined\manu
\begin{figure}
\else
\begin{figure*}[h!]
\fi
    \centerline{
        \scalebox{0.7}{
            \input{Fig/figQuantumCircuitSpectral3Qubits}
        }
    }
    \caption{Example of the quantum circuit with three qubits for calculating $U=\exp(-i\bD_\zeta^2)$.}
    \label{fig:qcfold}
\ifdefined\manu
\end{figure}
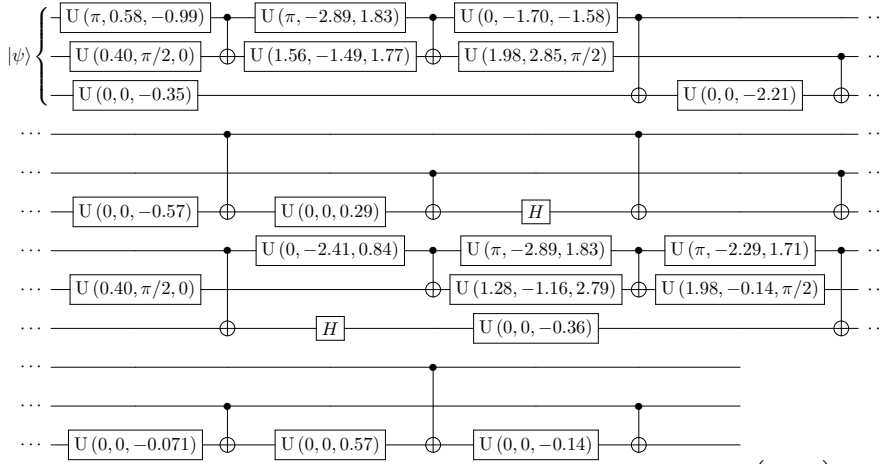
\else
\end{figure*}
\fi

\subsection{Algorithm complexity\label{subsec:complexity}} \addvspace{10pt}

The quantum SP method exploits the advantages of the efficient implementation of QFT and diagonal unitary operators.
The quantum FD method has $\lrN{\bA}_{\mr{max}}\sim \lrr{\lrn{u}/\Delta x + D/\Delta x^2 + \lrn{\alpha}}$. It is in the order of $N_t$ time steps in its classical counterpart.
In classical computing, the Courant-Friedreichs-Lewy condition~\cite{Courant1967} suggests $N_t = \mc{O}\lrr{2^{n_x}}$.
The complexity to perform the unitary operator $\exp(-i\bH t)$ in Eq.~\eqref{eq:evoFD} is then $\mc{O}(s\lrr{\bA}\lrr{n_x+n_p}2^{n_x+n_p})$.

Table~\ref{tab:complexity} summarizes the complexities of the quantum SP and FD methods, along with their classical counterparts for comparison.
Due to the efficiency of Hamiltonian simulations on a quantum computer, the quantum SP method achieves an exponential speedup compared with the classical one.
For the quantum FD method, the advantage of quantum computing is more clear for higher-dimensional problems, where the number of grid points grows as $N_x^d$ and the number of qubits rises as $dn_x$ with the number of dimension $d$.

\begin{table*}[h!] \footnotesize
    \caption{Algorithm complexities of the quantum/classical SP and FD methods.
    }
    \centering
    \begin{tabular}{llll}
    \toprule[1.5pt]
    Algorithm & Fourier transform & Time evolution & Total \\
    \midrule[1.0pt]
    Quantum SP      & $\mc{O}(n_x^2+n_p^2)$
                    & $\mc{O}((n_x+n_p)^2)$
                    & $\mc{O}(n_x^2+n_p^2+n_xn_p)$          \\
    Classical SP    & $\mc{O}(n_x 2^{n_x})$
                    & $\mc{O}(2^{2n_x})$
                    & $\mc{O}((n_x+2^{n_x}) 2^{n_x})$       \\
    Quantum FD      & $\mc{O}(n_p^2)$
                    & $\mc{O}(s(\bA)(n_x+n_p)2^{(n_x+n_p)})$
                    & $\mc{O}(s(\bA)(n_x+n_p)2^{(n_x+n_p)}+n_p^2)$ \\
    Classical FD    &
                    & $\mc{O}(s(\bA) 2^{2n_x})$
                    & $\mc{O}(s(\bA) 2^{2n_x})$ \\
    \bottomrule[1.5pt]
    \end{tabular}
\label{tab:complexity}
\end{table*}

\section{Results\label{sec:results}} \addvspace{10pt}



We validate the algorithms by employing IBM's Qiskit~\cite{Qiskit}, an open-source tool for the quantum computing simulator.
Qiskit enables simulating quantum computing on classical computers.
We firstly employ the ``StatevectorSimulator'' in Qiskit without noises to provide an ideal representation of the algorithm. 
Then we simulate with different noise models via the ``AerSimulator'' to assess the algorithm on the current and projected quantum computing simulators.

The central difference scheme is used for spatial discretization in both quantum and classical FD methods.
The third-order explicit Runge-Kutta~\cite{Bogacki1989} is employed for time marching in classical computing.
The source code used for the following cases is available online~\cite{qcRF}.

\subsection{Simulations without noise} \addvspace{10pt}


First, we test the quantum algorithm for 1D configuration with periodic boundary conditions.
To test the performance of different algorithms on evolving a scalar profile with multiple wavenumbers, we use the initial condition
\EQ
    \phi(t=0,x) = \sum_{k_s} \sin\lrr{k_s x} + \sum_{k_c} \cos\lrr{k_c x},
\EN
where $k_s$ and $k_c$ are wavenumbers for sine and cosine functions, respectively.
We set parameters $u=4$, $D=1$, and $\alpha=-0.2$, and wavenumbers $k_s=1,3$ and $k_c=2$.
With this initial condition, Eq.~\eqref{eq:linearPDE} has the analytical solution
\EQ\label{eq:exact}
\begin{split}
    \phi(t,x) = \sum_{k_s} \sin\lrs{k_s \lrr{x-ut}}e^{\lrr{-Dk_s^2+\alpha}t} \\
              + \sum_{k_c} \cos\lrs{k_c \lrr{x-ut}}e^{\lrr{-Dk_c^2+\alpha}t}.
\end{split}
\EN

\begin{figure}[h!]
    \centering
    \includegraphics[width=67 mm]{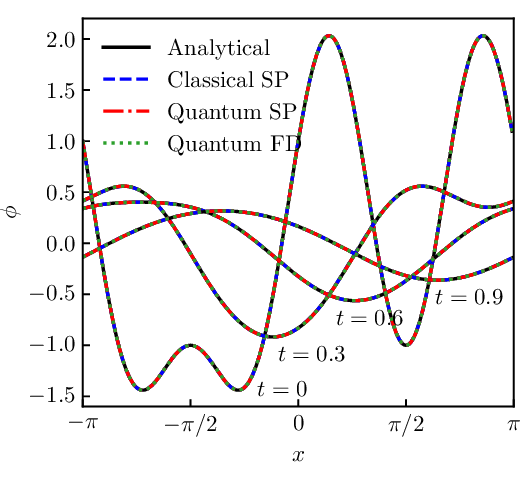}
    \caption{
        Comparison of analytical and numerical solutions (with the quantum SP, quantum FD, and classical SP methods) of $\phi\lrr{t,x}$ for the scalar evolution of a sinuous wave at different times.
    }
    \label{fig:Fourier}
\end{figure}

The results obtained via the quantum SP and FD methods are compared with the exact solutions and a classical SP simulation for validation.
The computational domain with $L_x=2\pi$ is discretized by $N_x=256$ points.
The quantum algorithm employs $n_x=8$ qubits for $x$, and $n_p=10$ qubits for the auxiliary dimension with $L_p=8\pi$.
The profiles of $\phi$ at $t=$0, 0.3, 0.6, and 0.9 in quantum and classical simulations, along with the exact solution, are compared in Fig.~\ref{fig:Fourier}.
All numerical methods perform well for this simple problem.

\begin{figure}[h!]
    \centering
    \includegraphics[width=67 mm]{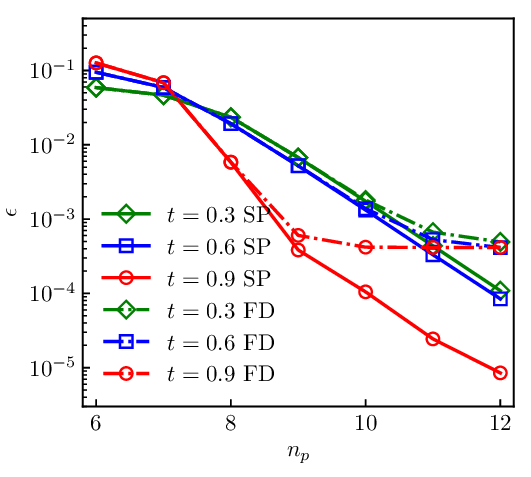}
    \caption{$L_2$ norm of the relative error by the quantum SP and FD methods against the number $n_p$ of qubits for the $p$-dimension.}
    \label{fig:ErrorNp}
\end{figure}

To quantify the accuracy of numerical results, we employ the $L_2$ norm of the relative error
\EQ
    \epsilon = \sqrt{\sum\lrn{\phi-\phi_A}^2} / \sqrt{\sum{\lrn{\phi_A}}^2},
\EN
where $\phi_A$ denotes the analytical solution in Eq.~\eqref{eq:exact}.
Figure~\ref{fig:ErrorNp} shows that variation of $\epsilon$ in the quantum simulation with different $n_p$ at $t=$0.3, 0.6, and 0.9.
Other numerical parameters including $n_x=8$ and $L_p=8\pi$ are fixed.
For comparison, $\epsilon$ obtained via the classical SP method are $1.76\times 10^{-3}$, $8.27\times 10^{-4}$, and $7.91\times 10^{-4}$ at $t=$ 0.3, 0.6, and 0.9, respectively.
%
The results show a second-order of accuracy with $\Delta p = L_p/2^{n_p}$.
The errors of moments (not shown) also confirm the effective accuracy control via $\Delta p$.
Regarding the accuracy and limit on the available qubits, we set $n_p=$ 9 or 10 with $L_p$ varying from $8\pi$ to $16\pi$ in the following cases.
The minimum $\epsilon$ in the quantum FD method depends on the spatial discretization scheme.

\begin{figure}[h!]
    \centering
    \includegraphics[width=67 mm]{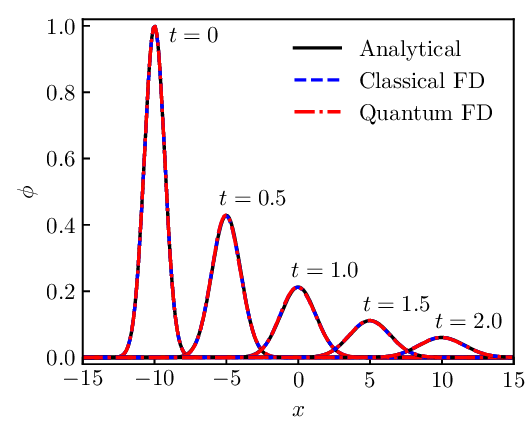}
    \caption{
        Comparison of analytical approximation and numerical solutions (with the quantum and classical FD methods) of $\phi\lrr{t,x}$ for the scalar evolution of a Gaussian wave at different times.
    }
    \label{fig:Gaussian}
\end{figure}

The simulation time is limited by $L_p/(2\lrN{\bm{a}}_\infty$), where $a_j$ is given by Eq.~\eqref{eq:speedSP} and the eigenvalue of $\bH_1$ for the quantum SP and FD methods, respectively.
Consequently, the quantum SP method may perform not well for problems with high wavenumbers.
We use the quantum FD method to simulate a convection-diffusion-reaction process of an initial Gaussian scalar function $\phi = \exp\lrs{-\lrr{x-\mu}^2}$ with $\mu=-10$.
We set parameters $u=10$, $D=0.5$, $\alpha=-1$, and $L_x=30$ for this case.
The number of grid points is $N_x=256$, equivalent to $n_x=8$ qubits for $x$.
When the scalar bump is remote from boundaries, the analytical approximation is
\EQ
    \phi = \dfrac{e^{\alpha t}}{\sqrt{1+4Dt}}\exp\lrs{-\dfrac{\lrr{x-\mu-ut}^2}{1+4Dt}}.
\EN
Figure~\ref{fig:Gaussian} compares quantum and classical FD results with the analytical approximation at $t=$0.5, 1.0, 1.5, 2.0.
The quantum simulation has $L_p=8\pi$ and $n_p=10$.
The results from the quantum simulation agree well with the classical numerical results and the analytical approximation.


We further validate the quantum algorithm for flows with inlet-outlet boundary conditions.
This configuration is often used in laminar~\cite{Kee2017} and turbulent~\cite{Domingo2023} flame simulations to investigate the flame speed and flame propagation.

In this validation case, we set the initial profile as the error function $\phi(0,x) = 0.5\lrs{\erf\lrr{x}+1}$.
The Dirichlet boundary condition $\phi=0$ and Neumann boundary condition $\partial\phi/\partial x=0$ are applied at the left inlet and right outlet, respectively.
Parameters are set as $u=5$, $D=0.01$, and $\alpha=-1$.
Computational domain has $L_x=30$, discretized with $256$ points, corresponding to $n_x=8$ qubits.
The quantum FD method sets $L_p=16\pi$ and $n_p=10$.

\begin{figure}[h!]
    \centering
    \includegraphics[width=67 mm]{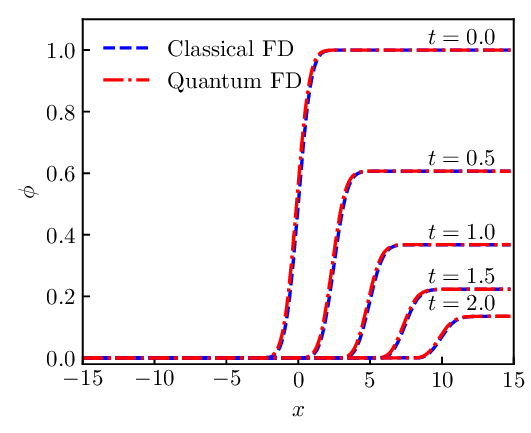}
    \caption{
        Comparison of results of $\phi\lrr{t,x}$ from the quantum and classical FD methods for the scalar evolution of a step wave at different times.
    }
    \label{fig:InOut1D}
\end{figure}

Figure~\ref{fig:InOut1D} shows the numerical results of the 1D inlet-outlet problem using the quantum and classical FD methods.
The scalar step profile moves towards the outlet due to convection, the transition region widens due to diffusion, and its overall magnitude decays due to the reaction source term.
The quantum result agrees well with the classical one.

\begin{figure}[h!]
    \centering
    \includegraphics[width=67 mm]{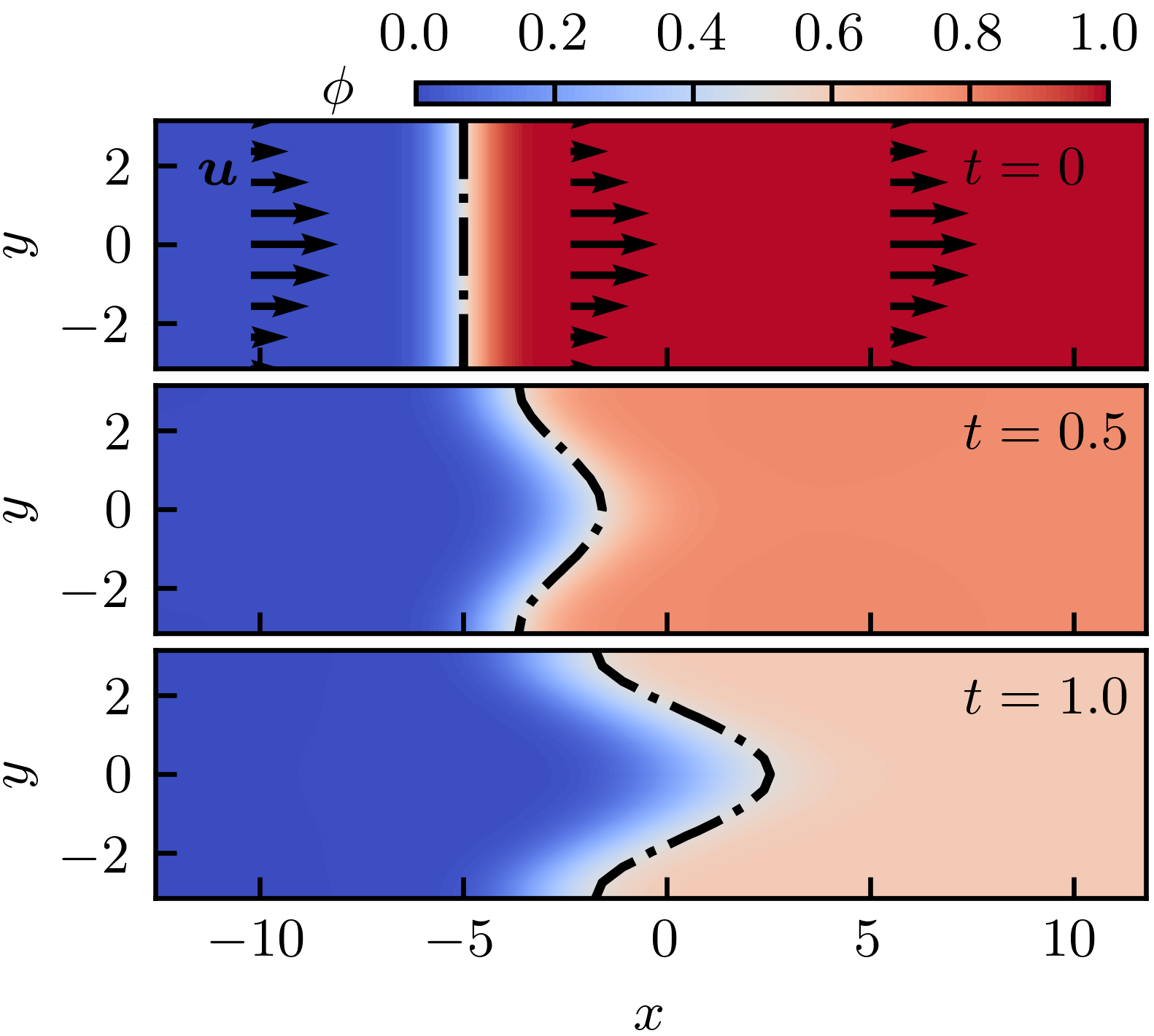}
    \caption{
        Comparison of results from the quantum (color contours) and classical (dash-dotted contour lines of $\phi=0.5$) FD methods for the evolution of  $\phi\lrr{t,x}$ in a 2D shear flow at different times. The arrows represent the velocity.
    }
    \label{fig:InOut2D}
\end{figure}

We further carry out a 2D validation of the quantum FD method.
The computational domain is a rectangle with $L_x=8\pi$ and $L_y=2\pi$, discretized on a $64\times 16$ grid, corresponding to 10 qubits. 
The $p$-dimension uses $n_p=9$ qubits with $L_p=16\pi$.
The boundary conditions $\phi=0$ and $\partial \phi/\partial x = 0$ are set at $x=-L_x/2$ and $x=L_x/2$, respectively.
Periodic boundary conditions are applied in the $y$-direction.
Parameters are $D=0.1$ and $\alpha=-0.5$.
The initial conditions are $\phi(0,x,y) = 0.5\lrs{\erf\lrr{x+5}+1}$, $u_x=2\cos\lrr{y}+4$ and $u_y=0$, illustrated in Fig.~\ref{fig:InOut2D}.

Figure~\ref{fig:InOut2D} plots the numerical results of the quantum and classical FD methods using the contour and the dash-dotted contour line of $\phi=0.5$, respectively.
Similar to the 1D case, the quantum FD method properly produces the convection, diffusion, and reaction processes.
The shear flow transports the scalar and causes a curved interface.
Therefore, the quantum simulations reproduce the almost identical results from classical simulations.

\subsection{Simulations with noises} \addvspace{10pt}

The noise level on real quantum computers varies based on the design and operation of the experiment facility. 
To elucidate the effects of noise in quantum computing, we carried out simulations using three different noise models, listed in Tab.~\ref{tab:noise}.
These noise models correspond to the noise levels of the current state, near-term target, and mid-term target.
The current state is referred by that used for quantum simulation of 2-D unsteady flows~\cite{Meng2024b}. It has the fidelities of 99.97\% and 99.83\% for the single-qubit and two-qubit gates, respectively. 
The near-term noise model reflects the cutting-edge noise level of a superconducting quantum processor~\cite{Li2023}.  
The single-qubit and two-qubit gate fidelities are 99.99\% and 99.90\%, respectively. 
The mid-term target is projected based on advancements expected in the next five years, with single-qubit and two-qubit gate fidelities of 99.999\% and 99.99\%, respectively. 

\begin{table}[h!] \footnotesize
    \caption{
    Noise models used in AerSimulator
    }
    \centering
    \begin{tabular}{lrr}
    \toprule[1.5pt]
    Gate fidelity               & Single-qubit  & Two-qubit     \\
    \midrule[1.0pt]
    Current~\cite{Meng2024b}    & 99.97\%       & 99.83\%       \\
    Near-term~\cite{Li2023}     & 99.99\%       & 99.90\%       \\
    Mid-term                    & 99.999\%      & 99.99\%       \\
    \bottomrule[1.5pt]
    \end{tabular}
\label{tab:noise}
\end{table}

\begin{figure}[h!]
    \centering
    \includegraphics[width=67 mm]{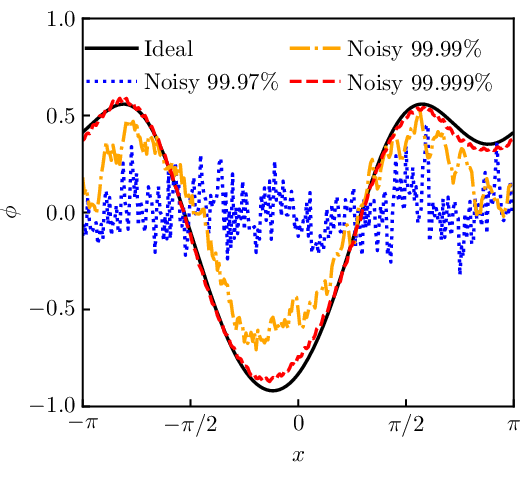}
    \caption{
        Comparison of the quantum spectral method simulation results using the ideal, current, near-term, and mid-term noise models for process in Eq.~\eqref{eq:exact} at $t=0.3$.
    }
    \label{fig:Noise}
\end{figure}

Figure~\ref{fig:Noise} compares the results of the quantum spectral method simulated using the ideal, current, near-term, and mid-term models for the convection-diffusion-reaction process in Eq.~\eqref{eq:exact} at $t=0.3$.
We observe that the current noise level introduces a significant error into the solution.
Simultaneously, the simulation with the near-term noise model gives a similar trend with the ideal solution.
Furthermore, a one-digit enhancement in the gate fidelity at the mid-term target improve accuracy significantly, yielding results closely aligned with the ideal simulation.

\section{Conclusions\label{sec:conclusions}} \addvspace{10pt}

We explore quantum computing of reacting flows by simulating Hamiltonian dynamics.
The warped phase transform~\cite{Jin2023c} is used to transform  the scalar transport equation into a Hamiltonian system.
In other words, the dissipative and non-Hermitian problem in physical space is mapped to a Hermitian one in a higher-dimensional space.
In the implementation, we develop the quantum SP and FD methods for simulating reacting flows in periodic and general conditions, respectively.
These methods provide a ``one-shot'' solution for a given time with no temporal discretization, avoiding frequent quantum state preparation and measurement.

Furthermore, we provide algorithms and quantum circuits for simulating the corresponding Hamiltonian dynamics using unitary operators.
The computational complexities of the quantum algorithms are compared to their classical counterparts.
The quantum SP method, which utilizes the QFT and diagonal unitary operators, exhibits exponential speedup over the classical SP method.
The quantum FD method can achieve exponential speedup in high-dimensional problems, although the complexities of quantum and classical FD methods are of the same order for 1D problems.

The validation of the quantum algorithms is conducted using the quantum computing simulator with Qiskit.
The quantum SP and FD methods are tested for several 1D and 2D flows with periodic or inlet-outlet boundary conditions.
The results obtained via ideal quantum computing agree well with the analytical and classical simulation results, capturing convection, diffusion, and reaction processes.
In addition, the error of quantum simulations is controlled by the auxiliary variable introduced in the warped phase transform.

We further assess the algorithm using the noisy quanutm computing simulators.
Despite the current noise level causing large errors, a single-digit improvement in the gate fidelities can significantly improve the performance of quantum computing.
These results demonstrate the potential of quantum computing as an effective and efficient method for simulating reacting flows.

In the future work, the major challenge for quantum computing of reacting flows is how to handle the nonlinear term within the linear Hamiltonian simulation framework~\cite{Liu2021,Jin2023a}.
In particular, the transported probability density function method~\cite{Pope1985,Haworth2010} can transform the nonlinear chemical source term into a linear convection term in the composition space.
Furthermore, coupling velocity and scalar remains challenging in the present framework of Hamiltonian simulation. 
It is anticipated to integrate the flow dynamics and scalar transport in quantum computing of reacting flows in the future work. 

\acknowledgement{Declaration of competing interest} \addvspace{10pt}

The authors declare no competing interests.




\acknowledgement{Acknowledgments} \addvspace{10pt}

The authors thank Z. Meng and J. Zhang for helpful suggestions.
This work has been supported in part by the National Natural Science Foundation of China (Nos.~52306126, 11925201, 11988102, and 92270203), the National Key R\&D Program of China (Grant No.~2020YFE0204200 and 2023YFB4502600), and the Xplore Prize.

%
%


 \footnotesize
 \baselineskip 9pt


\bibliographystyle{pci}
\bibliography{PCI_QC}


\newpage

\small
\baselineskip 10pt



\end{document}

%% file: Fig/figQuantumCircuitSpectralFull.tex
\begin{pgfpicture}{0em}{0em}{0em}{0em}
    \color{lime}
    \pgfrect[fill]{\pgfpoint{2.53em}{-6.15em}}{\pgfpoint{2.62em}{6.6em}}
    \pgfrect[fill]{\pgfpoint{14.72em}{-6.15em}}{\pgfpoint{3.em}{6.6em}}
    \color{orange}
    \pgfrect[fill]{\pgfpoint{7.16em}{-13.75em}}{\pgfpoint{5.54em}{14.2em}}
    \color{cyan}
    \pgfrect[fill]{\pgfpoint{2.53em}{-13.75em}}{\pgfpoint{2.62em}{6.6em}}
    \pgfrect[fill]{\pgfpoint{14.72em}{-13.75em}}{\pgfpoint{3.em}{6.6em}}
\end{pgfpicture}
\Qcircuit @C=1em @R=1em @!R {
    & \qw       & \multigate{3}{\mc{QF}}   
    & \qw       & \multigate{7}{U=e^{-iHt}}
    & \qw       & \multigate{3}{\mc{QF}^\dagger}   
    & \qw       & \qw
    \\
    & \qw       & \ghost{\mc{QF}}          
    & \qw       & \ghost{U=e^{-iHt}}
    & \qw       & \ghost{\mc{QF}^\dagger}          
    & \qw       & \qw
    \\
    & \vdots    & \nghost{\mc{QF}}         
    & \vdots    & \nghost{U=e^{-iHt}}
    & \vdots    & \nghost{\mc{QF}^\dagger}         
    & \vdots    &
    \\
    & \qw       & \ghost{\mc{QF}}          
    & \qw       & \ghost{U=e^{-iHt}}
    & \qw       & \ghost{\mc{QF}^\dagger}          
    & \qw       & \qw
    \gategroup{1}{9}{4}{9}{.5em}{\}}
    \inputgrouph{1}{4}{2.8em}{\ket{\phi(t)}}{-21em}
    \inputgroupv{1}{4}{0.5em}{2.8em}{\ket{\phi(0)}}
    \\
    & \qw       & \multigate{3}{\mc{QF}}   
    & \qw       & \ghost{U=e^{-iHt}}
    & \qw       & \multigate{3}{\mc{QF}^\dagger}   
    & \qw       & \rstick{\ket{1}} \qw
    \\
    & \qw       & \ghost{\mc{QF}}          
    & \qw       & \ghost{U=e^{-iHt}}
    & \qw       & \ghost{\mc{QF}^\dagger}          
    & \qw       & \rstick{\ket{0}} \qw
    \\
    & \vdots    & \nghost{\mc{QF}}         
    & \vdots    & \nghost{U=e^{-iHt}}
    & \vdots    & \nghost{\mc{QF}^\dagger}         
    & \vdots    & \quad\quad\vdots
    \\
    & \qw       & \ghost{\mc{QF}}          
    & \qw       & \ghost{U=e^{-iHt}}
    & \qw       & \ghost{\mc{QF}^\dagger}          
    & \qw       & \rstick{\ket{0}} \qw
    \inputgroupv{5}{8}{0.5em}{2.78em}{\ket{v}}
}

%% file: Fig/figQuantumCircuitSpectral3Qubits.tex
\Qcircuit @C=0.5em @R=0.5em @!R {
    \lstick{}
    & \gate{\mathrm{U}\,(\mathrm{\pi,0.58,-0.99})} 
    & \ctrl{1} 
    & \gate{\mathrm{U}\,(\mathrm{\pi,-2.89,1.83})} 
    & \ctrl{1} 
    & \gate{\mathrm{U}\,(\mathrm{0,-1.70,-1.58})} 
    & \ctrl{2} 
    & \qw
    & \qw
    & \rstick{\cdots} \qw
    \\
    \lstick{}
    & \gate{\mathrm{U}\,(\mathrm{0.40,\pi/2,0})} 
    & \targ 
    & \gate{\mathrm{U}\,(\mathrm{1.56,-1.49,1.77})} 
    & \targ 
    & \gate{\mathrm{U}\,(\mathrm{1.98,2.85,\pi/2})} 
    & \qw
    & \qw
    & \ctrl{1}
    & \rstick{\cdots} \qw
    \\
    \lstick{}
    & \gate{\mathrm{U}\,(\mathrm{0,0,-0.35})} 
    & \qw 
    & \qw 
    & \qw 
    & \qw 
    & \targ 
    & \gate{\mathrm{U}\,(\mathrm{0,0,-2.21})} 
    & \targ 
    & \rstick{\cdots} \qw
        \inputgroupv{1}{3}{1.0em}{2.1em}{\ket{\psi}}
    \\
    \lstick{\cdots} 
    & \qw
    & \ctrl{2} 
    & \qw 
    & \qw 
    & \qw 
    & \ctrl{2} 
    & \qw 
    & \qw 
    & \rstick{\cdots} \qw
    \\
    \lstick{\cdots} 
    & \qw
    & \qw 
    & \qw 
    & \ctrl{1} 
    & \qw 
    & \qw 
    & \qw 
    & \ctrl{1} 
    & \rstick{\cdots} \qw
    \\
    \lstick{\cdots} 
    & \gate{\mathrm{U}\,(\mathrm{0,0,-0.57})} 
    & \targ 
    & \gate{\mathrm{U}\,(\mathrm{0,0,0.29})} 
    & \targ 
    & \gate{H} 
    & \targ 
    & \qw
    & \targ 
    & \rstick{\cdots} \qw
    \\
    \lstick{\cdots} 
    & \qw 
    & \ctrl{2} 
    & \gate{\mathrm{U}\,(\mathrm{0,-2.41,0.84})} 
    & \ctrl{1} 
    & \gate{\mathrm{U}\,(\mathrm{\pi,-2.89,1.83})} 
    & \ctrl{1} 
    & \gate{\mathrm{U}\,(\mathrm{\pi,-2.29,1.71})} 
    & \ctrl{2} 
    & \rstick{\cdots} \qw
    \\
    \lstick{\cdots} 
    & \gate{\mathrm{U}\,(\mathrm{0.40,\pi/2,0})} 
    & \qw 
    & \qw 
    & \targ 
    & \gate{\mathrm{U}\,(\mathrm{1.28,-1.16,2.79})} 
    & \targ 
    & \gate{\mathrm{U}\,(\mathrm{1.98,-0.14,\pi/2})} 
    & \qw 
    & \rstick{\cdots} \qw
    \\
    \lstick{\cdots} 
    & \qw
    & \targ 
    & \gate{H} 
    & \qw 
    & \gate{\mathrm{U}\,(\mathrm{0,0,-0.36})} 
    & \qw 
    & \qw 
    & \targ 
    & \rstick{\cdots} \qw
    \\
    \lstick{\cdots} 
    & \qw & \qw & \qw 
    & \ctrl{2} 
    & \qw & \qw 
    & \qw
    \\
    \lstick{\cdots} 
    & \qw & \ctrl{1} 
    & \qw & \qw & \qw 
    & \ctrl{1} 
    & \qw
    \\
    \lstick{\cdots} 
    & \gate{\mathrm{U}\,(\mathrm{0,0,-0.071})} 
    & \targ & \gate{\mathrm{U}\,(\mathrm{0,0,0.57})} 
    & \targ & \gate{\mathrm{U}\,(\mathrm{0,0,-0.14})} 
    & \targ 
    & \qw
    \\
}

%% file: QCRF_arXiv_v3.bbl
\begin{thebibliography}{10}
\expandafter\ifx\csname url\endcsname\relax
  \def\url#1{\texttt{#1}}\fi
\expandafter\ifx\csname urlprefix\endcsname\relax\def\urlprefix{URL }\fi
\expandafter\ifx\csname href\endcsname\relax
  \def\href#1#2{#2} \def\path#1{#1}\fi

\bibitem{Nielson2010}
M.~A. Nielsen, I.~L. Chuang, Quantum Computation and Quantum Information: 10th
  Anniversary Edition, Cambridge University Press, Cambridge, 2010.

\bibitem{Feynman1982}
R.~P. Feynman, Simulating physics with computers, Int. J. Theor. Phys. 21
  (1982) 467--488.

\bibitem{Lloyd1996}
S.~Lloyd, Universal quantum simulators, Science 273 (1996) 1073--1078.

\bibitem{Low2017}
G.~H. Low, I.~L. Chuang, Optimal {Hamiltonian} simulation by quantum signal
  processing, Phys. Rev. Lett. 118 (2017) 010501.

\bibitem{Childs2018}
A.~M. Childs, D.~Maslov, Y.~Nam, N.~J. Ross, Y.~Su, Toward the first quantum
  simulation with quantum speedup, Proc. Natl. Acad. Sci. 115 (2018)
  9456--9461.

\bibitem{Campbell2019}
E.~Campbell, Random compiler for fast {Hamiltonian} simulation, Phys. Rev.
  Lett. 123 (2019) 070503.

\bibitem{Givi2020}
P.~Givi, A.~J. Daley, D.~Mavriplis, M.~Malik, Quantum speedup for aeroscience
  and engineering, AIAA J. 58 (2020) 3715--3727.

\bibitem{Xu2019}
G.~Xu, A.~J. Daley, P.~Givi, R.~D. Somma, Quantum algorithm for the computation
  of the reactant conversion rate in homogeneous turbulence, Combust. Theory
  Model. 23 (2019) 1090--1104.

\bibitem{Akiba2023}
T.~Akiba, Y.~Morii, K.~Maruta, Carleman linearization approach for chemical
  kinetics integration toward quantum computation, Sci. Rep. 13 (2023) 3935.

\bibitem{Domingo2023}
P.~Domingo, L.~Vervisch, Recent developments in {DNS} of turbulent combustion,
  Proc. Combust. Inst. 39 (2023) 2055--2076.

\bibitem{Lu2021}
Z.~Lu, Y.~Yang, Modeling pressure effects on the turbulent burning velocity for
  lean hydrogen/air premixed combustion, Proc. Combust. Inst. 38 (2021)
  2901--2908.

\bibitem{Su2022}
Y.~Su, Z.~Lu, Y.~Yang, Suppression of the turbulent kinetic energy and
  enhancement of the flame-normal {Reynolds} stress in premixed jet flames at
  small {Lewis} numbers, Combust. Flame 246 (2022) 112461.

\bibitem{Harrow2009}
A.~W. Harrow, A.~Hassidim, S.~Lloyd, Quantum algorithm for linear systems of
  equations, Phys. Rev. Lett. 103 (2009) 150502.

\bibitem{Childs2021}
A.~M. Childs, J.-P. Liu, A.~Ostrander, High-precision quantum algorithms for
  partial differential equations, Quantum 5 (2021) 574.

\bibitem{Clader2013}
B.~D. Clader, B.~C. Jacobs, C.~R. Sprouse, Preconditioned quantum linear system
  algorithm, Phys. Rev. Lett. 110 (2013) 250504.

\bibitem{Krovi2023}
H.~Krovi, Improved quantum algorithms for linear and nonlinear differential
  equations, Quantum 7 (2023) 913.

\bibitem{Costa2019}
P.~C.~S. Costa, S.~Jordan, A.~Ostrander, Quantum algorithm for simulating the
  wave equation, Phys. Rev. A 99 (2019) 012323.

\bibitem{Wei2023}
S.-J. Wei, C.~Wei, P.~Lv, C.~Shao, P.~Gao, Z.~Zhou, K.~Li, T.~Xin, G.-L. Long,
  A quantum algorithm for heat conduction with symmetrization, Sci. Bull. 68
  (2023) 494--502.

\bibitem{Liu2023}
B.~Liu, L.~Zhu, Z.~Yang, G.~He, Quantum implementation of numerical methods for
  convection-diffusion equations: Toward computational fluid dynamics, Commun.
  Comput. Phys. 33 (2023) 425--451.

\bibitem{Hoefler2023}
T.~Hoefler, T.~H\"{a}ner, M.~Troyer, Disentangling hype from practicality: On
  realistically achieving quantum advantage, Commun. ACM 66 (2023) 82--87.

\bibitem{Meng2023}
Z.~Meng, Y.~Yang, Quantum computing of fluid dynamics using the hydrodynamic
  {Schr\"odinger} equation, Phys. Rev. Res. 5 (2023) 033182.

\bibitem{Yang2023}
Y.~Yang, S.~Xiong, Z.~Lu, Applications of the vortex-surface field to flow
  visualization, modelling and simulation, Flow 3 (2023) E33.

\bibitem{Meng2024a}
Z.~Meng, Y.~Yang, Quantum spin representation for the {Navier-Stokes} equation
  (2024).
\newblock \href {http://arxiv.org/abs/2403.00596} {\path{arXiv:2403.00596}}.

\bibitem{An2023}
D.~An, J.-P. Liu, D.~Wang, Q.~Zhao, A theory of quantum differential equation
  solvers: limitations and fast-forwarding (2023).
\newblock \href {http://arxiv.org/abs/2211.05246} {\path{arXiv:2211.05246}}.

\bibitem{Jin2023c}
S.~Jin, N.~Liu, Y.~Yu, Quantum simulation of partial differential equations:
  applications and detailed analysis, Phys. Rev. A 108 (2023) 032603.

\bibitem{Preskill2018}
J.~Preskill, Quantum computing in the {NISQ} era and beyond, Quantum 2 (2018)
  79.

\bibitem{Daley2022}
A.~J. Daley, I.~Bloch, C.~Kokail, S.~Flannigan, N.~Pearson, M.~Troyer,
  P.~Zoller, Practical quantum advantage in quantum simulation, Nature 607
  (2022) 667--676.

\bibitem{Qiskit}
{Qiskit contributors}, Qiskit: An open-source framework for quantum computing
  (2023).

\bibitem{Liu2021}
J.-P. Liu, H.~O. Kolden, H.~K. Krovi, N.~F. Loureiro, K.~Trivisa, A.~M. Childs,
  Efficient quantum algorithm for dissipative nonlinear differential equations,
  Proc. Natl. Acad. Sci. 118 (2021) e2026805118.

\bibitem{Joseph2020}
I.~Joseph, {Koopman}--von {Neumann} approach to quantum simulation of nonlinear
  classical dynamics, Phys. Rev. Res. 2 (2020) 043102.

\bibitem{Jin2023a}
S.~Jin, N.~Liu, Y.~Yu, Time complexity analysis of quantum algorithms via
  linear representations for nonlinear ordinary and partial differential
  equations, J. Comput. Phys. 487 (2023) 112149.

\bibitem{Weinstein2001}
Y.~S. Weinstein, M.~A. Pravia, E.~M. Fortunato, S.~Lloyd, D.~G. Cory,
  Implementation of the quantum {Fourier} transform, Phys. Rev. Lett. 86 (2001)
  1889--1891.

\bibitem{Meng2024b}
Z.~Meng, J.~Zhong, S.~Xu, K.~Wang, J.~Chen, F.~Jin, X.~Zhu, Y.~Gao, Y.~Wu,
  C.~Zhang, N.~Wang, Y.~Zou, A.~Zhang, Z.~Cui, F.~Shen, Z.~Bao, Z.~Zhu, Z.~Tan,
  T.~Li, P.~Zhang, S.~Xiong, H.~Li, Q.~Guo, Z.~Wang, C.~Song, H.~Wang, Y.~Yang,
  Simulating unsteady fluid flows on a superconducting quantum processor
  (2024).
\newblock \href {http://arxiv.org/abs/2404.15878} {\path{arXiv:2404.15878}}.

\bibitem{Courant1967}
R.~Courant, K.~Friedrichs, H.~Lewy, On the partial difference equations of
  mathematical physics, IBM J. Res. Dev. 11 (1967) 215--234.

\bibitem{Bogacki1989}
P.~Bogacki, L.~Shampine, A 3(2) pair of {Runge}--{Kutta} formulas, Appl. Math.
  Lett. 2 (1989) 321--325.

\bibitem{qcRF}
Z.~Lu, Y.~Yang, {qcReactingFlows}, \url{github.com/YYgroup/qcReactingFlows}
  (2024).

\bibitem{Kee2017}
R.~J. Kee, M.~E. Coltrin, P.~Glarborg, H.~Zhu, Chemically Reacting Flow, John
  Wiley \& Sons, Ltd, Hoboken, NJ, 2017.

\bibitem{Li2023}
Z.~Li, P.~Liu, P.~Zhao, Z.~Mi, H.~Xu, X.~Liang, T.~Su, W.~Sun, G.~Xue, J.-N.
  Zhang, W.~Liu, Y.~Jin, H.~Yu, Error per single-qubit gate below 10$^{-4}$ in
  a superconducting qubit, npj Quantum Inf. 9 (2023) 111.

\bibitem{Pope1985}
S.~B. Pope, {PDF} methods for turbulent reactive flows, Prog. Energy Combust.
  Sci. 11 (1985) 119--192.

\bibitem{Haworth2010}
D.~C. Haworth, Progress in probability density function methods for turbulent
  reacting flows, Prog. Energy Combust. Sci. 36 (2010) 168--259.

\end{thebibliography}
